\documentclass[11pt,a4paper]{article}

\usepackage{epsfig,amsmath,amssymb,cite}

\begin{document}

\title{Stability of Chaplygin gas thin--shell wormholes} 
\author{Ernesto F. Eiroa$^{1,}$\thanks{e-mail: eiroa@iafe.uba.ar}, 
Claudio Simeone$^{2,}$\thanks{e-mail: csimeone@df.uba.ar}\\
{\small $^1$ Instituto de Astronom\'{\i}a y F\'{\i}sica del Espacio, C.C. 67, 
Suc. 28, 1428, Buenos Aires, Argentina}\\
{\small $^2$ Departamento de F\'{\i}sica, Facultad de Ciencias Exactas y 
Naturales,} \\ 
{\small Universidad de Buenos Aires, Ciudad Universitaria Pab. I, 1428, 
Buenos Aires, Argentina}} 

\maketitle

\begin{abstract}
In this paper we construct spherical thin--shell wormholes supported by a Chaplygin gas. For a rather general class of geometries we introduce a new approach for the stability analysis of static solutions under perturbations preserving the symmetry. We apply this to wormholes constructed from  Schwarzschild, Schwarzschild--de Sitter, Schwarzschild--anti de Sitter and Reissner--Nordstr\"{o}m metrics. In the last two cases, we find that there are values of the parameters for which stable static solutions exist.\\

\noindent 
PACS number(s): 04.20.Gz, 04.40.Nr, 98.80.Jk

\end{abstract}

\section{Introduction}\label{intro} 

Traversable Lorentzian wormholes \cite{motho} are solutions of the equations of gravitation associated to a nontrivial topology of the spacetime: their basic feature is that they connect  two regions  (of the same universe or two separate universes \cite{motho, visser}) by a throat. For the case of static wormholes the throat is defined as a minimal area surface satisfying a flare-out condition \cite{ hovis1}. To fulfill this, wormholes must be threaded by exotic matter that violates the null energy condition \cite{motho, visser, hovis1, hovis2};  it was shown by Visser \textit{et al} \cite{viskardad}, however, that the amount of exotic matter needed around the throat can be made as small as desired by means of an  appropriate choice of  the geometry of the wormhole.  \\

A well studied class of wormholes is that of thin--shell ones, which are constructed by cutting and pasting two manifolds \cite{visser, mvis} to form a geodesically complete new one with a shell placed  in the joining surface. This makes such wormholes of particular interest because the exotic matter needed for the existence of the configuration is located only at the shell. Stability analysis of thin--shell wormholes under perturbations preserving the original symmetries has been widely developed.  A linearized  analysis  of a thin--shell wormhole made by joining two Schwarzschild geometries was performed  by Poisson and Visser in Ref. \cite{poisson}. Later, the same method was applied to  wormholes constructed using branes with negative tensions in Ref. \cite{barcelo}, and  the case
of transparent spherically symmetric thin--shells and wormholes was studied in Ref. \cite{ishak}. The linearized stability analysis was extended to Reissner--Nordstr\"{o}m thin--shell geometries in Ref.  \cite{eirom}, and to wormholes with a cosmological constant in Ref. \cite{lobo}. The case of dynamical thin--shell wormholes was considered in Ref. \cite{lobo2}. The stability and energy conditions for five dimensional thin--shell wormholes in Einstein--Maxwell theory with a Gauss--Bonnet term were studied in Ref. \cite{marc}, while thin--shell wormholes associated with cosmic strings have been treated in Refs. \cite{eisi}. Other related works can be found in Refs. \cite{other}.\\

The requirement of matter violating energy conditions relates the study of wormholes to modern cosmology:  Current day observational data seem to point towards an accelerated expansion of the universe \cite{acc}. If General Relativity is assumed as the right gravity theory describing the large scale behavior of the universe, this implies that its energy density $\rho$ and pressure $p$ should violate the strong energy condition.
Several models for the matter leading to such  situation have been proposed \cite{matt}. One of them is the Chaplygin gas \cite{chap}, a perfect fluid fulfilling  the equation of state
$p\rho =-A$, where $A$ is a positive constant. A remarkable property of the Chaplygin gas is that the squared sound velocity $v_{s}^{2}=A/\rho^{2}$ is always positive even in the case of exotic matter. Though introduced for purely phenomenological reasons (in fact, not related with cosmology \cite{aero}), such an equation of state has the interesting feature of being derivable from string theory; more precisely, it can be obtained from the Nambu--Goto action for $d-$branes moving in $(d+2)-$dimensional spacetime if one works in the light-cone parametrization \cite{brane}. Besides, an analogous equation of state, but with $A$ a negative constant, was introduced for describing cosmic strings with small structure (``wiggly'' strings)  \cite{wiggly}. \\
    
Models of exotic matter of interest in cosmology have already  been considered in wormhole construction. Wormholes supported by ``phantom energy'' (with equation of state $p=\omega \rho, \ \omega<-1$) have been studied in detail \cite{phantom}.
A generalized Chaplygin gas, with equation of state $p\rho^\alpha=-A$ $(0<\alpha\leq 1)$, has been  proposed by   Lobo in Ref. \cite{lobo73} as the exotic matter supporting a wormhole of the Morris--Thorne type \cite{motho}; there,  as a possible way to keep the exotic matter  within a finite region of space,  matching the wormhole metric to an exterior vacuum metric was proposed. If, instead, a thin--shell wormhole is constructed, exotic matter can be  restricted from the beginning to the shell located at the joining surface. In the present paper we study  spherically symmetric thin--shell wormholes with matter in the form of the Chaplygin gas (the generalized Claplygin gas introduces a new constant and non trivial complications in the equations, which possibly cannot be solved in an analytical way). We introduce a new approach for the study of the stability under radial perturbations. The more complex case of the stability analysis of thin--shell wormholes under perturbations that do not preserve the symmetry has not been addressed in previous works, even for simpler metrics and equations of state, so we consider it beyond the scope of this article. In Section \ref{tswh} we apply the Darmois--Israel formalism to the cut and paste construction of a generic wormhole with  the Chaplygin equation of state imposed on the matter of the shell. In Section \ref{stab} we perform a detailed analysis of the stability under spherically symmetric perturbations. In Section \ref{exa} we analyze the specific cases of the Schwarzschild and the Reissner--Nordstr\"{om} geometries, and we also consider  the inclusion of  a cosmological constant of arbitrary sign. In Section \ref{conclu} the results are discussed. We adopt units such that $c=G=1$.\\

\section{Wormhole construction}\label{tswh}

Let us consider a spherically symmetric metric of the form
\begin{equation} 
ds^2=-f(r)dt^2+f(r)^{-1}dr^2+r^2 (d\theta ^2+\sin^2\theta d\varphi^2), 
\label{e1}
\end{equation}
where $r>0$ is the radial coordinate, $0\le \theta \le \pi$ and $0\le \varphi<2\pi $ are the angular coordinates, and $f(r)$ is a positive function from a given radius. For the construction of the thin--shell wormholes, we choose a radius $a$, take two identical copies of the region with $r\geq a$:
\begin{equation} 
\mathcal{M}^{\pm }=\{X^{\alpha }=(t,r,\theta,\varphi)/r\geq a\},  \label{e2}
\end{equation}
and paste them at the hypersurface
\begin{equation} 
\Sigma \equiv \Sigma ^{\pm }=\{X/F(r)=r-a=0\},  \label{e3}
\end{equation}
to create a new manifold $\mathcal{M}=\mathcal{M}^{+}\cup \mathcal{M}^{-}$. If the metric (\ref{e1}) has an event horizon with radius $r_{h}$, the value of $a$ should be greater than $r_{h}$, to avoid the presence of horizons and singularities. This construction produces a geodesically complete manifold, which has two regions connected by a throat with radius $a$, where the surface of minimal area is located and the condition of flare-out is satisfied. On this manifold we can define a new radial coordinate $l=\pm \int_{a}^{r}\sqrt{1/f(r)}dr$ representing the proper radial distance to the throat, which is situated at $l=0$; the plus and minus signs correspond, respectively, to $\mathcal{M}^{+}$ and $\mathcal{M}^{-}$. We follow the standard Darmois-Israel formalism \cite{daris,mus} for its study and we let the throat radius $a$ be a function of time. The wormhole throat $\Sigma $ is a synchronous timelike hypersurface, where  we  define coordinates $\xi ^{i}=(\tau ,\theta,\varphi )$, with $\tau $ the proper time on the shell.  The second fundamental forms (extrinsic curvature) associated with the two sides of the shell are:
\begin{equation} 
K_{ij}^{\pm }=-n_{\gamma }^{\pm }\left. \left( \frac{\partial ^{2}X^{\gamma
} } {\partial \xi ^{i}\partial \xi ^{j}}+\Gamma _{\alpha \beta }^{\gamma }
\frac{ \partial X^{\alpha }}{\partial \xi ^{i}}\frac{\partial X^{\beta }}{
\partial \xi ^{j}}\right) \right| _{\Sigma },  \label{e4}
\end{equation}
where $n_{\gamma }^{\pm }$ are the unit normals ($n^{\gamma }n_{\gamma }=1$) to $\Sigma $ in $\mathcal{M}$:
\begin{equation} 
n_{\gamma }^{\pm }=\pm \left| g^{\alpha \beta }\frac{\partial F}{\partial
X^{\alpha }}\frac{\partial F}{\partial X^{\beta }}\right| ^{-1/2}
\frac{\partial F}{\partial X^{\gamma }}.  \label{e5}
\end{equation}
Working in the orthonormal basis $\{e_{\hat{\tau}},e_{\hat{\theta}},
e_{\hat{\varphi}}\}$ ($e_{\hat{\tau}}=e_{\tau }$, 
$e_{\hat{\theta}}=a^{-1}e_{\theta }$, 
$e_{\hat{\varphi}}=(a\sin \theta )^{-1}e_{\varphi }$), for the metric (\ref{e1}) we have that
\begin{equation} 
K_{\hat{\theta}\hat{\theta}}^{\pm }=K_{\hat{\varphi}\hat{\varphi}}^{\pm
}=\pm \frac{1}{a}\sqrt{f(a)+\dot{a}^2},
\label{e6}
\end{equation}
and
\begin{equation} 
K_{\hat{\tau}\hat{\tau}}^{\pm }=\mp \frac{f'(a)+2\ddot{a}}{2\sqrt{f(a)+\dot{a}^2}} ,
\label{e7}
\end{equation}
where a prime and the dot stand for the derivatives with respect to $r$ and $\tau$, respectively. Defining $[K_{_{\hat{\imath}\hat{\jmath}}}]\equiv K_{_{\hat{\imath}\hat{\jmath}
}}^{+}-K_{_{\hat{\imath}\hat{\jmath}}}^{-}$, $K=tr[K_{\hat{\imath}\hat{
\jmath }}]=[K_{\; \hat{\imath}}^{\hat{\imath}}]$ and introducing  the surface stress-energy tensor $S_{_{\hat{\imath}\hat{\jmath} }}={\rm diag}(\sigma ,p_{\hat{\theta}},p_{\hat{\varphi}})$ we obtain the Einstein equations on the shell (the Lanczos equations):
\begin{equation} 
-[K_{\hat{\imath}\hat{\jmath}}]+Kg_{\hat{\imath}\hat{\jmath}}=8\pi 
S_{\hat{\imath}\hat{\jmath}},\label{e8}
\end{equation}
which in our case correspond  to a shell of radius $a$ with energy density $\sigma$ and transverse pressure $p=p_{\hat{\theta}}=p_{\hat{\varphi}}$ given by
\begin{equation} 
\sigma=-\frac{1}{2\pi a}\sqrt{f(a)+\dot{a}^2},
\label{e9}
\end{equation}
\begin{equation} 
p=\frac{1}{8\pi a}\frac{2a\ddot{a}+2\dot{a}^2+2f(a)+af'(a)}{\sqrt{f(a)+\dot{a}^2}}.
\label{e10}
\end{equation}
The equation of state for a Chaplygin gas has the form
\begin{equation}
p=\frac{-A}{\sigma},
\label{e11} 
\end{equation} 
where $A$ is a positive constant. Replacing Eqs. (\ref{e9}) and (\ref{e10}) in Eq. (\ref{e11}), we obtain 
\begin{equation}
2a\ddot{a}+2\dot{a}^2-16\pi ^2Aa^2+2f(a)+af'(a)=0.
\label{e12} 
\end{equation} 
This is the differential equation that should be satisfied by the throat radius of thin--shell wormholes threaded by exotic matter with the equation of state of a Chaplygin gas.

\section{Stability of static solutions}\label{stab}

From Eq. (\ref{e12}), the static solutions, if they exist, have a throat radius $a_{0}$ that should fulfill the equation
\begin{equation}
-16\pi ^2Aa_{0}^2+2f(a_{0})+a_{0}f'(a_{0})=0,
\label{e13} 
\end{equation} 
with the condition $a_{0}>r_{h}$ if the original metric has an event horizon. The surface energy density and pressure are given in the static case by
\begin{equation} 
\sigma=-\frac{\sqrt{f(a_{0})}}{2\pi a_{0}},
\label{e14}
\end{equation}
and
\begin{equation} 
p=\frac{2\pi Aa_{0}}{\sqrt{f(a_{0})}}.
\label{e15}
\end{equation}
The existence of static solutions depends on the explicit form of the function $f$. To study the stability of the static solutions under perturbations preserving the symmetry it is convenient to rewrite $a(\tau )$ in the form
\begin{equation}
a(\tau )=a_{0}[1+\epsilon (\tau )],
\label{e16} 
\end{equation} 
with $\epsilon (\tau )\ll 1$ a small perturbation. Replacing Eq. (\ref{e16}) in Eq. (\ref{e12}) and using Eq. (\ref{e13}), we obtain
\begin{equation}
(1+\epsilon)\ddot{\epsilon}+\dot{\epsilon}^2-8\pi^2A(2+\epsilon)\epsilon+g(a_{0},\epsilon)=0,
\label{e17} 
\end{equation} 
where the function $g$ is defined by
\begin{equation}
g(a_{0},\epsilon)=\frac{2f(a_{0}+a_{0}\epsilon)+a_{0}(1+\epsilon)f'(a_{0}+a_{0}\epsilon)}
{2a_{0}^2}-\frac{2f(a_{0})+a_{0}f'(a_{0})}{2a_{0}^2}.
\label{e18} 
\end{equation} 
Defining $\nu (\tau)= \dot{\epsilon }(\tau)$, Eq. (\ref{e17}) can be written as a set of first order differential equations
\begin{eqnarray}
\dot{\epsilon} & = & \nu \nonumber \\
\dot{\nu } & = & \frac{8\pi^2A(2+\epsilon)\epsilon-g(a_{0},\epsilon)}{1+\epsilon}-\frac{\nu ^2}{1+\epsilon}.
\label{e19} 
\end{eqnarray} 
Taylor expanding to first order in $\epsilon $ and $\nu $ we have
\begin{eqnarray}
\dot{\epsilon} & = & \nu \nonumber \\
\dot{\nu } & = & \Delta \epsilon,
\label{e20} 
\end{eqnarray} 
where
\begin{equation}
\Delta=16\pi ^2A-\frac{\partial g}{\partial \epsilon}(a_{0},0)=16\pi ^2 A-\frac{3f'(a_{0})+a_{0}f''(a_{0})}{2a_{0}},
\label{e21} 
\end{equation} 
which, by defining
\begin{equation}
\xi=\left(\begin{array}{c}
\epsilon \\ 
\nu
\end{array}\right)
\;\; \mathrm{and}\;\; 
M=\left(\begin{array}{cc}
0 & 1 \\ 
\Delta & 0
\end{array}\right),
\end{equation} 
can be put in the matrix form
\begin{equation}
\dot{\xi}=M\xi.
\label{e23} 
\end{equation} 
If $\Delta>0$ the matrix $M$ has two real eigenvalues: $\lambda_{1}=-\sqrt{\Delta}<0$ and $\lambda_{2}=\sqrt{\Delta}>0$. The presence of an eigenvalue with positive real part makes this case unstable. As the imaginary parts of the eigenvalues are zero, the instability is of saddle type. When $\Delta=0$, we have $\lambda_{1}=\lambda_{2}=0$, and to first order in $\epsilon$ and $\nu$ we obtain $\nu=constant=\nu_{0}$ and $\epsilon=\epsilon_{0}+\nu_{0}(\tau-\tau_{0})$, so the static solution is unstable. If $\Delta<0$ there are two imaginary eigenvalues $\lambda_{1}=-i\sqrt{|\Delta |}$ and $\lambda_{2}=i\sqrt{|\Delta |}$; in this case the linear system does not determine the stability and the set of nonlinear differential equations should be taken into account. For the analysis of the $\Delta<0$ case, we can rewrite Eq. (\ref{e19}) in polar coordinates $(\rho, \gamma )$, with $\epsilon=\rho \cos \gamma $ and $\nu=\rho \sin \gamma $, and make a first order Taylor expansion in $\rho$, which gives
\begin{eqnarray}
\dot{\rho} &=& \sin \gamma \cos \gamma (1+\Delta)\rho \nonumber \\
\dot{\gamma} &=& \Delta\cos^2\gamma -\sin^2\gamma +h(\gamma )\rho ,
\label{e25} 
\end{eqnarray} 
where $h(\gamma )$ is a bounded periodic function of $\gamma $. For small values of $\rho$, i.e. close to the equilibrium point, the time derivative of the angle $\gamma $ is negative (the leading term $\Delta \cos^2\gamma -\sin^2 \gamma $ is negative), then $\gamma $ is a monotonous decreasing function of time, so the solution curves rotate clockwise around the equilibrium point. To see that these solution curves are closed orbits for small $\rho$, we take a time $\tau_{1}$ so that $(\epsilon(\tau_{1}),\nu(\tau_{1}))=(\epsilon_{1},0)$ with $\epsilon_{1}>0$. As the solution curve passing through $(\epsilon_{1},0)$ rotates clockwise around $(0,0)$, there will be a time $\tau_{2}>\tau_{1}$ such that the curve will cross the $\epsilon$ axis again, in the point $(\epsilon(\tau_{2}),\nu(\tau_{2}))=(\epsilon_{2},0)$, with $\epsilon_{2}<0$. As the Eq. (\ref{e19}) is invariant under the transformation composed of a time inversion $\tau\rightarrow-\tau$ and the inversion $\nu\rightarrow-\nu$, the counterclockwise curve beginning in $(\epsilon_{1},0)$ should cross the $\epsilon$ axis also in $(\epsilon_{2},0)$. Therefore, for $\Delta<0$ the solution curves of Eq. (\ref{e19}) should be closed orbits near the equilibrium point $(0,0)$, which is a stable center. The only stable static solutions with throat radius $a_{0}$ are then those which have $\Delta<0$, and they are not asymptotically stable, i.e. when perturbed the throat radius oscillates periodically around the equilibrium radius, without settling down again.

\section{Application to different geometries}\label{exa}

In this section we analyze wormholes constructed with different metrics with the form of Eq. (\ref{e1}).

\subsection{Schwarzschild case}\label{schw}

\begin{figure}[t!]
\begin{center}
\vspace{0cm}
\includegraphics[width=11cm]{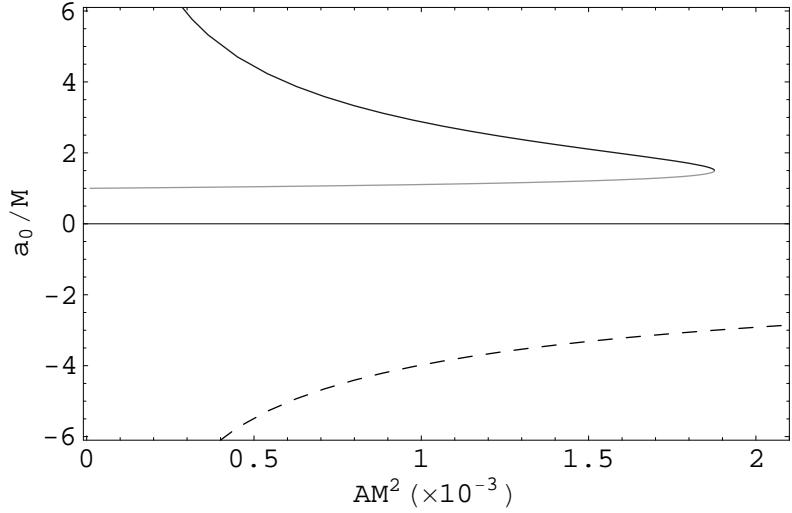}
\vspace{0cm}
\end{center} 
\caption{Solutions of Eq. (\ref{s4}) as functions of $AM^2$. For $AM^2<(54\pi ^2)^{-1}$ there are three real roots, two of them positive $a_{0}^{u}$ (full black line) and  $a_{0}^{s}$  (full grey line), and one negative $a_{0}^{neg}$ (dashed line); when $AM^2=(54\pi ^2)^{-1}$ the two positive roots merge into a double one; and if $AM^2>(54\pi ^2)^{-1}$ there is only one negative real root.}
\label{fig1}
\end{figure}

For the Schwarzschild metric we have that
\begin{equation}
f(r)=1-\frac{2M}{r},
\label{s1} 
\end{equation} 
where $M$ is the mass. This geometry has an event horizon situated at $r_h=2M$, so the radius of the wormhole throat should be taken greater than $2M$. The surface energy density and pressure for static solutions are given by
\begin{equation}
\sigma=-\frac{\sqrt{a_{0}-2M}}{2\pi a_{0}^{3/2}},
\label{s2} 
\end{equation} 
and
\begin{equation} 
p=\frac{2\pi Aa_{0}^{3/2}}{\sqrt{a_{0}-2M}},
\label{s3}
\end{equation} 
with the throat radius $a_{0}$ that should satisfy the cubic equation:
\begin{equation}
8\pi ^2Aa_{0}^3-a_{0}+M=0.
\label{s4} 
\end{equation}
Using that $A>0$ and $M>0$, it is not difficult to see that this equation has one negative real root and two non real roots if $AM^2>(54\pi ^2)^{-1}$, one negative real root and one double positive real root if $AM^2=(54\pi ^2)^{-1}$, and one negative real root and two positive real roots if $AM^2<(54\pi ^2)^{-1}$. The solutions of cubic equation (\ref{s4}), plotted in Fig. \ref{fig1}, are given by\footnote{the notation for the roots will be clear below}
\begin{equation}
a_{0}^{neg}=\frac{-1-i\sqrt{3}-\left(1-i \sqrt{3} \right) \left(-3 \pi \sqrt{6A} M+ i \sqrt{1-54 \pi ^{2} A M^{2}} \right)^{2/3}}{4 \pi \sqrt{6 A}\left(-3\pi \sqrt{6 A} M+i \sqrt{1-54 {{\pi }^2} A M^{2} } \right)^{1/3}},
\label{s5} 
\end{equation} 
\begin{equation}
a_{0}^{s}=\frac{-1+ i \sqrt{3}- \left(1+ i \sqrt{3} \right) \left(-3  \pi \sqrt{6 A} M+ i \sqrt{1-54 \pi ^{2} A M^{2}} \right)^{2/3}}{4 \pi \sqrt{6 A} \left(-3 \pi \sqrt{6 A} M+ i \sqrt{1-54 \pi ^{2} A {M^2}} \right)^{1/3}},
\label{s6} 
\end{equation} 
and
\begin{equation}
a_{0}^{u}=\frac{1+\left(-3 \pi \sqrt{6A} M+ i \sqrt{1-54 \pi ^{2} A M^{2}} \right)^{2/3}}
{2 \pi \sqrt{6A} \left(-3 \pi \sqrt{6A}  M+ i \sqrt{1-54 \pi ^{2}A M^{2}}\right)^{1/3}},
\label{s7} 
\end{equation} 
where the powers of complex numbers give the principal value. The negative root $a_{0}^{neg}$ has no physical meaning, so if $AM^2>(54\pi ^2)^{-1}$ there are no static solutions. When $AM^2=(54\pi ^2)^{-1}$ the positive double real solution of Eq. (\ref{s4}) is $a_{0}=3M/2<r_{h}$, then no static solutions are present. For $AM^2<(54\pi ^2)^{-1}$, the positive roots of Eq. (\ref{s4}) are $M<a_{0}^{s}\le 3M/2$ and $3M/2\le a_{0}^{u}$; the first one is always smaller than $r_{h}$, thus it has to be discarded, and the second one is greater than $r_{h}$ if $AM^2<\alpha _{0}$, where $\alpha _{0}\approx 1.583 \times 10^{-3}<(54\pi ^2)^{-1}$ (obtained numerically). To study the stability of $a_{0}^{u}$, we calculate $\Delta $ for this case:
\begin{equation}
\Delta=16\pi ^2 A-\frac{M}{a_{0}^3},
\label{s8} 
\end{equation} 
which, with the help of Eq. (\ref{s4}), can be simplified to give:
\begin{equation}
\Delta=\frac{1}{a_{0}^3}(2a_{0}-3M).
\label{s9} 
\end{equation} 
Then, using that $a_{0}^{u}>3M/2$, it is easy to see that $\Delta $ is always a positive number so, following Sec. \ref{stab}, the static solution is unstable (saddle equilibrium point). Briefly, for the Schwarzschild metric if $AM^2\ge \alpha _{0}$ no static solutions are present, and if $AM^2<\alpha _{0}$ there is only one unstable static solution with throat radius $a_{0}^{u}$.

\subsection{Schwarzschild--de Sitter case}\label{schw-ds}

\begin{figure}[t!]
\begin{center}
\vspace{0cm}
\includegraphics[width=11cm]{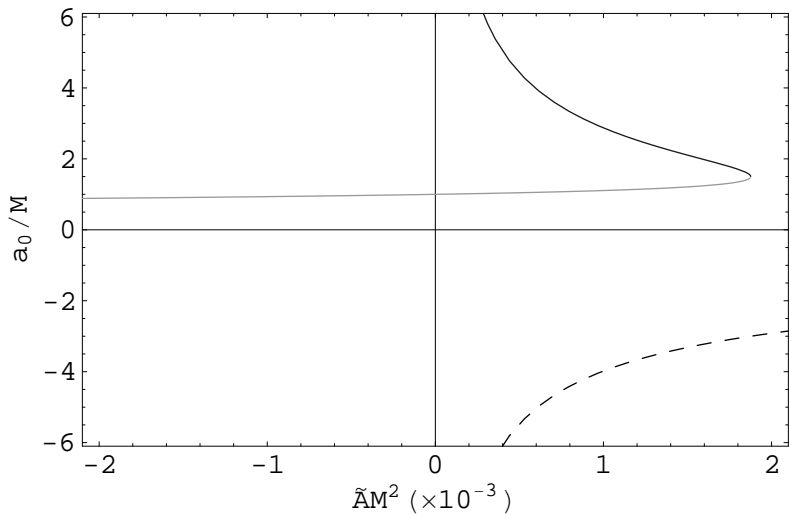}
\vspace{0cm}
\end{center} 
\caption{Solutions of Eq. (\ref{ds6}) as functions of $\tilde{A}M^2$, where $\tilde{A}=A+\Lambda /(12\pi ^2)$. We have that $\tilde{A}$ is always positive or it has any sign, depending on $\Lambda>0$ or $\Lambda<0$, respectively. When $\tilde{A}\le 0$, there is only one real (and positive) root $a_{0}^{s}$  (full grey line); for $0<\tilde{A}M^2<(54\pi ^2)^{-1}$ there are three real roots, two of them positive $a_{0}^{u}$ (full black line) and  $a_{0}^{s}$  (full grey line), and one negative $a_{0}^{neg}$ (dashed line); when $\tilde{A}M^2=(54\pi ^2)^{-1}$ the two positive roots merge into a double one; and if $\tilde{A}M^2>(54\pi ^2)^{-1}$ there is only one negative real root. }
\label{fig2}
\end{figure}

For the Schwarzschild--de Sitter metric the function $f$ has the form
\begin{equation}
f(r)=1-\frac{2M}{r}-\frac{\Lambda}{3}r^2,
\label{ds1} 
\end{equation} 
where $\Lambda>0$ is the cosmological constant. If $\Lambda M^{2}>1/9$ we have that $f(r)$ is always negative, so we take $0<\Lambda M^{2}\le 1/9$. In this case the geometry has two horizons, the event and the cosmological ones, which are placed, respectively, at
\begin{equation}
r_{h}=\frac{-1+i\sqrt{3}-(1+i\sqrt{3})\left(-3\sqrt{\Lambda}M+i\sqrt{1-9\Lambda M^{2}}
\right)^{2/3}}{2\sqrt{\Lambda}\left(-3\sqrt{\Lambda}M+i\sqrt{1-9\Lambda M^{2}}\right)^{1/3}},
\label{ds2} 
\end{equation} 
\begin{equation}
r_{c}=\frac{1+\left(-3\sqrt{\Lambda}M+i\sqrt{1-9\Lambda M^{2}}\right)^{2/3}}
{\sqrt{\Lambda}\left(-3\sqrt{\Lambda}M+i\sqrt{1-9\Lambda M^{2}}\right)^{1/3}}.
\label{ds3} 
\end{equation}
The event horizon radius $r_{h}$ is a continuous and increasing function of $\Lambda $, with  $r_{h}(\Lambda \rightarrow 0^{+})=2M$ and $r_{h}(\Lambda M^{2}=1/9)=3M$, and the cosmological horizon radius $r_{c}$ is a continuous and decreasing function of $\Lambda $, with $r_{c}(\Lambda \rightarrow 0^{+})\rightarrow +\infty$ and $r_{c}(\Lambda M^{2}=1/9)=3M$. If $0<\Lambda M^{2}<1/9$ the wormhole throat radius should be taken in the range $r_{h}<a_{0}<r_{c}$, and if $\Lambda M^{2}=1/9$ the construction of the wormhole is not possible, because $r_{h}=r_{c}=3M$. Using Eqs. (\ref{e14}) and (\ref{e15}), we obtain that the energy density and the pressure at the throat are given by
\begin{equation}
\sigma=-\frac{\sqrt{-\Lambda a_{0}^{3}+3a_{0}-6M}}{2\sqrt{3}\pi a_{0}^{3/2}},
\label{ds4} 
\end{equation} 
\begin{equation}
p=\frac{2\sqrt{3}\pi A a_{0}^{3/2}}{\sqrt{-\Lambda a_{0}^{3}+3a_{0}-6M}}.
\label{ds5} 
\end{equation} 
The throat radius $a_{0}$ should satisfy in this case the cubic equation
\begin{equation}
8\pi ^2\left( A+\frac{\Lambda}{12\pi ^2}\right) a_{0}^3-a_{0}+M=0.
\label{ds6} 
\end{equation}
If we define $\tilde{A}=A+\Lambda /(12\pi ^2)$, which in this case is a positive number, it is easy to see that Eq. (\ref{ds6}) has the same form as Eq. (\ref{s4}), so the solutions of Eq. (\ref{ds6}) are given again by Eqs. (\ref{s5}-\ref{s7}), with $A$ replaced by $\tilde{A}$. These solutions are shown in Fig. \ref{fig2} (the part of the plot with $\tilde{A}>0$). With the same arguments of Sec. \ref{schw}, if $\tilde{A}M^{2}\ge (54\pi ^{2})^{-1}$ we have no static solutions. Using that $2M<r_{h}<3M$, when $\tilde{A}M^{2}< (54\pi ^{2})^{-1}$, $a_{0}^{s}$ is always smaller than $r_{h}$, so it has to be discarded, and $a_{0}^{u}$ can be greater, equal or smaller than $r_{h}$, depending on the values of the parameters. To study the stability of the static solution with throat radius $a_{0}^{u}$ (if present) we obtain for this metric that $\Delta$ is again given by Eq. (\ref{s9}). Using that $a_{0}^{u}>3M/2$, it is straightforward to see that $\Delta $ is a positive number, therefore, following Sec. \ref{stab}, the static solution is unstable (saddle equilibrium point).

\subsection{Schwarzschild--anti de Sitter case}\label{schw-ads}

\begin{figure}[t!]
\begin{center}
\vspace{0cm}
\includegraphics[width=11cm]{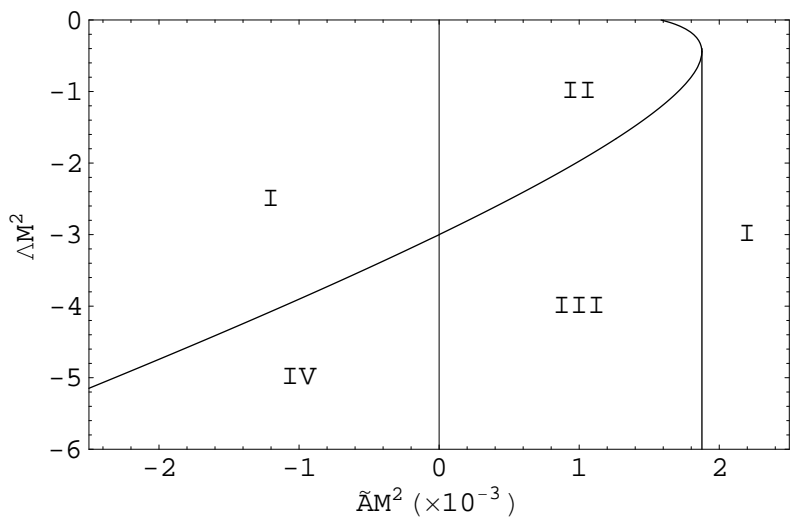}
\vspace{0cm}
\end{center} 
\caption{Schwarzschild--anti de Sitter case: number of static solutions for given parameters $M$, $\Lambda <0$ and $\tilde{A}=A+\Lambda /(12\pi ^{2})$. Region I: no static solutions; region II: one static unstable solution; region III: two static solutions, one stable and the other unstable; region IV: one static stable solution.}
\label{fig3}
\end{figure} 

In the Schwarzschild--anti de Sitter metric the function $f(r)$ is given again by Eq. (\ref{ds1}), but now with a negative $\Lambda $. The event horizon for this metric is placed at 
\begin{equation}
r_{h}=\frac{1-\left(-3\sqrt{|\Lambda|}M+\sqrt{1+9|\Lambda|M^{2}}\right)^{2/3}}
{\sqrt{|\Lambda|}\left(-3\sqrt{|\Lambda|}M+\sqrt{1+9|\Lambda|M^{2}}\right)^{1/3}}.
\label{ads1} 
\end{equation}
The horizon radius $r_{h}$ is a continuous and increasing function of $\Lambda$, with values in the interval $0<r_{h}<2M$, with $r_{h}(\Lambda\rightarrow -\infty)=0$ and $r_{h}(\Lambda \rightarrow 0^{-})=2M$. The wormhole throat radius $a_{0}$ should be greater than $r_{h}$. The energy density and the pressure at the throat are given by Eqs. (\ref{ds4}) and (\ref{ds5}), with the throat radius $a_{0}$ satisfying Eq. (\ref{ds6}). Using  $\tilde{A}=A+\Lambda /(12\pi ^2)$, which now can be positive, zero or negative, we have that Eq. (\ref{ds6}) takes again the form of Eq. (\ref{s4}) with $A$ replaced by $\tilde{A}$. Its solutions are shown in Fig. \ref{fig2}. For the stability analysis, it is not difficult to see that $\Delta$ is again given by Eq. (\ref{s9}). Then we have five possible situations, depending on the different values of $\tilde{A}$:
\begin{enumerate}
\item When $\tilde{A}M^{2}>(54\pi ^{2})^{-1}$ there are no positive real solutions of Eq. (\ref{ds6}), so we have no static solutions. 
\item When $\tilde{A}M^{2}=(54\pi ^{2})^{-1}$, Eq. (\ref{ds6}) has one positive double real root $a^{u}_{0}=3M/2$ for which $\Delta =0$, then following Sec. \ref{stab} it is unstable. For the existence of the unstable static solution a large enough value of $|\Lambda |$ is needed, so that $a_{0}^{u}>r_{h}$.
\item When $0<\tilde{A}M^{2}<(54\pi ^{2})^{-1}$, the solutions of Eq. (\ref{ds6}) are given by Eqs. (\ref{s5}-\ref{s7}), with $A$ replaced by $\tilde{A}$ (see Fig. \ref{fig2}).  Using that $a_{0}^{u}>3M/2$ and $M<a_{0}^{s}<3M/2$, it follows from Sec. \ref{stab} that the static solution with throat radius  $a_{0}^{u}$ is unstable (saddle equilibrium point) and the solution with radius $a_{0}^{s}$ is stable (center). This stable static solution exists if $|\Lambda |$ is large enough so that $a_{0}^{s}>r_{h}$.
\item When $\tilde{A}M^{2}=0$, Eq. (\ref{ds6}) has only one real solution given by $a^{s}_{0}=M$. The associated wormhole solution exists if $|\Lambda |$ is large enough so that $M>r_{h}$ and in this case it is stable (center) because $\Delta$ is negative.
\item When $\tilde{A}M^{2}<0$, we have that Eq. (\ref{ds6}) has only one real solution given by
\begin{equation}
a_{0}^{s}=\frac{1-{{ \left(-3 \pi  {\sqrt{6|\tilde{A}|}} M+{\sqrt{1+54 {{\pi }^2} |\tilde{A}| {M^2}}} \right)}^{2/3}}}{2  \pi {\sqrt{6|\tilde{A}|}} {{ \left(-3 \pi  {\sqrt{6|\tilde{A}|}}  M+{\sqrt{1+54 {{\pi }^2}|\tilde{A}| {M^2}}}\right)}^{1/3}}},
\label{ads}
\end{equation}
which is an increasing function of $\tilde{A}$ and lies in the range $0<a_{0}^{s}<M$ (see Fig. \ref{fig2}). Then we have that $\Delta<0$ and the solution is stable. Again, $|\Lambda |$ should be large enough to have $a_{0}^{s}>r_{h}$. 
\end{enumerate} 

The number of static solutions for the different values of the parameters is shown in Fig. \ref{fig3}. When $\tilde{A}M^{2}<(54\pi ^{2})^{-1}$ there are static stable solutions if $|\Lambda |$ is large enough so the condition of $a^{s}_{0}>r_{h}$ is satisfied, which corresponds to the regions III and IV of Fig. \ref{fig3}.

\subsection{Reissner--Nordstr\"{o}m case}\label{rn}

The Reissner--Norsdtr\"om metric represents a charged object with spherical symmetry which has 
\begin{equation}
f(r)=1-\frac{2M}{r}+\frac{Q^2}{r^2},
\label{rn1} 
\end{equation} 
where $Q$ is the charge. For $|Q|<M$ this geometry has an inner and an outer (event) horizon given by
\begin{equation}
r^{\pm}=M\pm \sqrt{M^{2}-Q^{2}},
\label{rn2} 
\end{equation} 
if $|Q|=M$ the two horizons merge into one, and when $|Q|>M$ there are no horizons and the metric represents a naked singularity. When $|Q|\le M$ the throat radius $a_{0}$ should be taken greater than $r_{h}=r^{+}$ so that no horizons are present in $\mathcal{M}$. If $|Q|>M$ the condition $a_{0}>0$ assures that the naked singularity is removed. Replacing Eq. (\ref{rn1}) in Eqs. (\ref{e14}) and (\ref{e15}), we obtain the energy density and pressure at the throat:
\begin{equation}
\sigma=-\frac{\sqrt{a_{0}^{2}-2Ma_{0}+{Q^2}}}{2\pi a_{0}^{2}},
\label{rn3} 
\end{equation} 
\begin{equation}
p=\frac{2\pi A a_{0}^{2}}{\sqrt{a_{0}^{2}-2Ma_{0}+{Q^2}}}.
\label{rn4} 
\end{equation}
\begin{figure}[t!]
\begin{center}
\vspace{0cm}
\includegraphics[width=11cm]{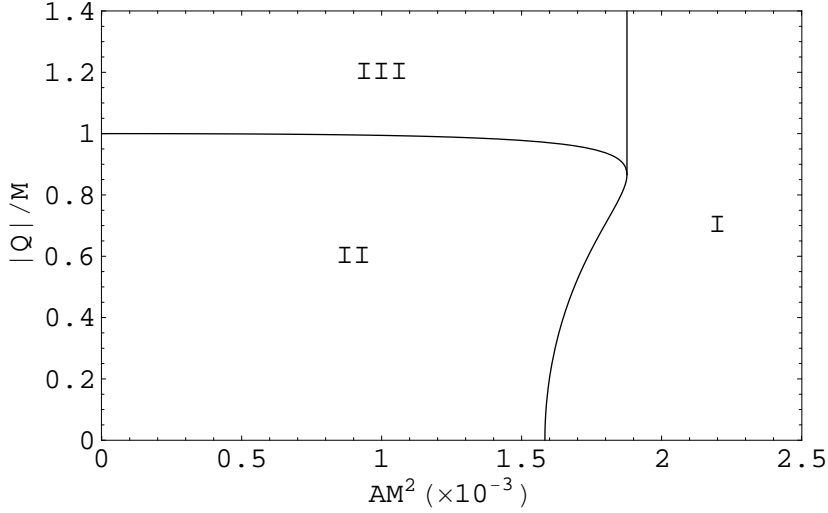}
\vspace{0cm}
\end{center} 
\caption{Reissner--Nordstr\"{o}m case: number of static solutions for given parameters $A$, $M$ and $Q$. Region I: no static solutions; region II: one static unstable solution; region III: two static solutions, one stable and the other unstable.}
\label{fig4}
\end{figure} 
Replacing the metric (\ref{rn1}) in Eq. (\ref{e13}), it is not difficult to see that the charge cancels out and the throat radius should satisfy the cubic equation (\ref{s4}) again,  with its solutions given by Eqs. (\ref{s5}-\ref{s7}) and plotted in Fig. \ref{fig1}. As pointed out in Sec. \ref{schw} the number of roots of the cubic (\ref{s4}) are zero, one or two depending on if $AM^2$ is, respectively,  greater, equal or smaller than $(54\pi ^2)^{-1}$. The solutions should satisfy that $a_{0}>r_{h}$, then the number of static solutions will also depend on the value of the charge. As $r_{h}$ is a decreasing function of $|Q|$, for large values of charge there will be two static solutions with radius $a_{0}^{s}$ and $a_{0}^{u}$. Following Sec. \ref{stab},  the relevant quantity for the analysis of the stability is $\Delta$, which is again given by Eq. (\ref{s9}). Using that when $AM^2<(54\pi ^2)^{-1}$ we have $a_{0}^{s}<3M/2$ and $a_{0}^{u}>3M/2$, it is easy to check that $\Delta<0$ for $a_{0}^{s}$ and $\Delta>0$ for $a_{0}^{u}$, therefore $a_{0}^{s}$ is stable (center) and $a_{0}^{u}$ is unstable (saddle). The special case where $AM^2=(54\pi ^2)^{-1}$ and $a_{0}^{u}=a_{0}^{s}=3M/2$ is also unstable because $\Delta=0$. The number of static solutions depends on the values of the parameters $A$, $M$ and $Q$. Using that $a_{0}^{u}$ and $a_{0}^{s}$ should be greater than $r_{h}$, given by Eq. (\ref{rn2}), and defining the functions
\begin{equation}
\alpha=\frac{1+ \left(-3 \pi \sqrt{6A} M+i\sqrt{1-54 \pi ^2 A {M^2}} \right)^{2/3}}{2  \pi \sqrt{6A}M \left(-3 \pi  \sqrt{6A} M+i\sqrt{1-54 \pi ^2 A M^2}\right)^{1/3}},
\end{equation}
and
\begin{equation}
\beta=\frac{-1+i\sqrt{3}-\left(1+ i \sqrt{3} \right) \left(-3 \pi \sqrt{6 A} M+ i \sqrt{1-54 \pi ^2 A M^2} \right)^{2/3}}{4 \pi \sqrt{6 A} M \left(-3 \pi \sqrt{6 A} M+i\sqrt{1-54 \pi ^2 A M^2} \right)^{1/3}},
\end{equation} 
it is easy to see that the three possible cases are:
\begin{figure}[t!]
\begin{center}
\vspace{0cm}
\includegraphics[width=11cm]{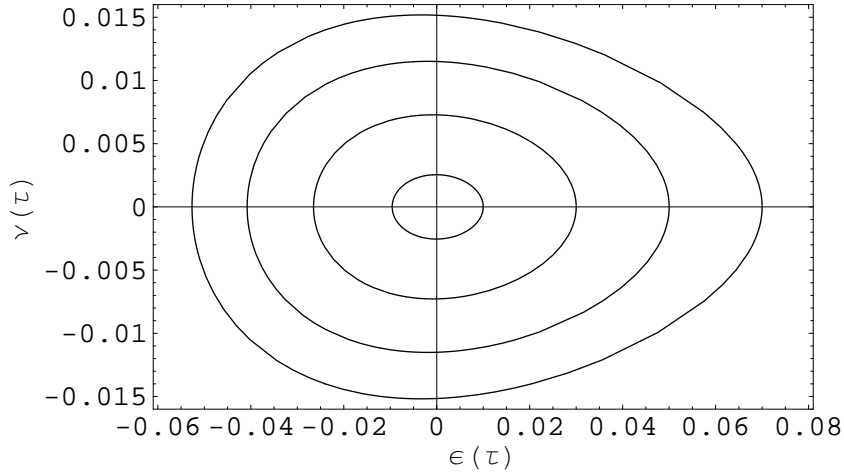}
\vspace{0cm}
\end{center} 
\caption{Reissner--Nordstr\"{o}m case: phase portrait around a stable static solution. The values of the parameters used in the plot are $|Q|/M=0.95$ and $AM^2=1.85\times10^{-3}$ and the radius of the throat is $a_{0}^{s}/M=1.406$. The perturbation $\epsilon (\tau )$ is defined by $\epsilon (\tau )=a(\tau )/a_{0}^{s}-1$  and $\nu (\tau )=\dot{\epsilon} (\tau )$ is its time derivative. The curves have different initial conditions and rotate clockwise.}
\label{fig5}
\end{figure}
\begin{enumerate}
\item No static solutions: when $AM^2>(54\pi ^2)^{-1}$, or if 
$\alpha_{0}<AM^{2}\le(54\pi ^2)^{-1}$ and $|Q|/M\le \sqrt{\alpha(2-\alpha)}$, where $\alpha_{0}\approx 1.583\times 10^{-3}$.
\item One unstable static solution: when $0\le AM^{2}<\alpha_{0}$ and $|Q|/M<\sqrt{\beta(2-\beta)}$, or if $\alpha_{0}\le AM^{2}\le (54\pi ^2)^{-1}$ and 
$\sqrt{\alpha(2-\alpha)}<|Q|/M\le \sqrt{\beta(2-\beta)}$.
\item Two static solutions, one stable and the other unstable: when 
$0\le AM^{2}<(54\pi ^2)^{-1}$ and $|Q|/M>\sqrt{\beta(2-\beta)}$. 
\end{enumerate}
The three regions are plotted in Fig. \ref{fig4}. For $|Q|>M$ and $AM^{2}<(54\pi ^2)^{-1}$ there is always one stable static solution (center). Also one stable static solution can be obtained with values of $|Q|$ smaller than $M$ when $AM^{2}$ is slightly smaller than $(54\pi ^2)^{-1}$. These stable configurations correspond to values of the parameters within the region III of Fig. \ref{fig4}. A phase portrait of curves surrounding a static stable solution is shown in Fig. \ref{fig5}.

\section{Conclusions}\label{conclu}

We have constructed spherical thin--shell wormholes supported by exotic matter fulfilling the Chaplygin gas equation of state. Such kind of exotic matter has been recently considered of particular interest in cosmology as it provides a possible explanation for the observed accelerated expansion of the Universe. For the wormhole construction we have applied the usual cut and paste procedure at a radius greater than the event horizon (if it exists) of each metric. Then, by considering the throat radius as a function of time we have obtained a general equation of motion for the Chaplygin gas shell. We have addressed the issue of stability of static configurations under perturbations preserving the symmetry. The procedure developed has been applied to wormholes constructed from Schwarzschild, Schwarzschild--de Sitter, Schwarzschild--anti de Sitter and Reissner--Nordstr\"{o}m geometries. In the pure Schwarzschild case we have found that no stable static configurations exist. A similar result has been obtained for the case of Schwarzschild--de Sitter metric (positive cosmological constant). For the Schwarzschild--anti de Sitter geometry, we have found that the existence of static solutions requires that the mass $M$,  the negative cosmological constant $\Lambda$ and the positive constant $A$ characterizing the Chaplygin fluid should satisfy  $\tilde{A} M^{2} \leq (54\pi^{2})^{-1}$, with $\tilde{A}=A+\Lambda/(12\pi ^{2})$, and  $|\Lambda |$ great enough to yield a small horizon radius in the original manifold. If these conditions are verified, for each combination of the parameters, when $0 <\tilde{A} M^{2}<(54\pi^{2})^{-1}$ there is one stable solution with throat radius $a_{0}$ in the range $M<a_{0}<3M/2$, and for $\tilde{A} M^{2}\le 0$ there is one stable configuration with $a_{0}\le M$. In the Reissner--Nordstr\"{o}m case the existence of static solutions with charge $Q$ requires  $AM^{2}\leq (54\pi^{2})^{-1}$; we have found that if this condition is satisfied, then one stable configuration always exists if $|Q|/M>1$. When $|Q|/M$ is slightly smaller than one, there is also a stable static solution if $AM^{2}$ is close to $(54\pi^{2})^{-1}$. In this work, then, we have shown that if over-densities in the Chaplygin cosmological fluid had taken place, stable static configurations which represent traversable wormholes would be possible.

\section*{Acknowledgments}

This work has been supported by Universidad de Buenos Aires and CONICET. Some calculations 
in this paper were done with the help of the package GRTensorII (which can be obtained freely at the address http://grtensor.org).

\end{document}